\documentstyle[12pt,oldlfont]{article}
\normalbaselines
\begin{document}
\bibliographystyle{unsrt}

\begin{center}
{\Large {\bf Deformation of Partical Distribution Functions
due to Q-nonlinearity and Nonstationary Casimir Effect}}
\end{center}

\begin{center}
V. I. Man'ko\\\ {\em Lebedev Physics Institute, Leninsky pr., 53\\
Moscow 117924, Russia\\}

\end{center}

\begin{abstract}The geometrical phase is shown to be integral of motion.
Deformation of particle distribution function corresponding to nonstationary
Casimir effect is expressed in terms of multivariable Hermite polynomials.
Correction to Planck distribution due to q--nonlinearity is discussed.
\end{abstract}

\section{Geometrical Phase as Integral of Motion}
The periodical in time quantum systems are described by the
quasi--energy states \cite{zel67}, \cite{rit67}. The notion of generalized
quasienergies for the nonstationary quantum systems with nonperiodical
Hamiltonians has been introduced in \cite{man92}. The main point
of the quasienergy concept is to relate the quasienrgies to the
eigenvalues of the Floquet operator which is equal to the evolution
operator of a quantum system taken at a given time moment. In \cite{man92}
the Floquet operator was related to integrals of motion and a new operator
was introduced which is the integral of motion and has the same quasienergy
spectrum that the Floquet operator. Also this concept has been applied
to Berry phase introduced in \cite{ber84}. We will review this result.

If one has the system with hermitian Hamiltonian $~H(t)$ such that
$~H(t+T)=H(t)$ the unitary evolution operator $~U(t)$ is defined as follows
\begin{equation}
|\psi,t>=U(t)|\psi,0>,
\end{equation}
where $~|\psi,0>$ is a state vector of the system taken at the initial
time moment. Then by definition the operator $~U(T)$, which is the analog
of the Floquet operator for periodical systems, has eigenvalues of the
form
\begin{equation}
f=\exp(-i\varphi T),
\end{equation}
where $~\varphi $ is called the phase containing the Berry phase. The
spectra of the phases may be either discrete or continuous ones (or
mixed) for different quantum systems. The operator $~U(T)$ does not
satisfy the relation
\begin{equation}
dI(t)/dt+i[H(t),I(t)]=0~~~~(\hbar=1),
\end{equation}
which defines the integral of motion $~I(t)$. So, the operator
$~U(T)$ is not the integral of motion for the nonstationary quantum
systems. But as it was pointed out in Ref.\cite{mal79} any operator of
the form
\begin{equation}
I(t)=U(t)I(0)U^{-1}(t)
\end{equation}
satisfies Eq. (3) and this operator is the integral of motion for
the quantum system under study. Let us apply this ansatz to the case of
nonstationary systems whose parameters are the same after time $~T$. We
introduce the unitary operator $~M(t)$ which has the form
\begin{equation}
M(t)=U(t)U(T)U^{-1}(t).
\end{equation}
This operator is the integral of motion due to the construction given
by Eq. (4) for any integral of motion. The spectrum of the new invariant
operator $~M(t)$ coincides with the spectrum of the operator
$~U(T)$. We have proved that since geometrical phases are defined as
eigenvalues of the integral of motion $~M(t)$ they are conserved
quantities.

The operator (5) may be represented in the form
\begin{equation}
M(t)=\exp[-iTH_{ef}(t)],
\end{equation}
where $~H_{ef}(t)$ is the invariant effective Hamiltonian. Its spectrum
provides the geometrical phase of the system.

The quasienergy spectrum of periodically kicked quantum systems may
be connected with quantum chaos phenomenon (see Refs. \cite{chi79},
\cite{haa92}, \cite{kar91}, \cite{zhi94}. In the Ref.\cite{haa92} the
integral of motion for delta--kicked nonlinear oscillator has been found
to exist even in the case of chaotic behaviour.In the Ref.\cite{kar91}
the symmetry group criterium for the periodically delta-kicked systems
has been found to obtain either regular or chaotic behaviour of these
systems. The criterium relates the Floquet operator spectrum to the
conjugacy class of the system symmetry group. For the quadratic hermitian
Hamiltonians this group turnes out to be the real symplectic group
$~Sp(2N,R)$. The results of the Ref.\cite{kar91} may be applied to
the systems with Berry phase, too. The geometrical phase
is determined by the conjugacy class of the same group to which belongs
the integral of motion (5). For quadratic systems it is the same real
symplectic group $~Sp(2N,R)$.

If the Hamiltonian of the nonstationary system has the symmetry property
\begin{equation}
H(t+iT)=H(t),
\end{equation}
where $~T$ is a real number the loss-energy states exist as analogs of
quasienergy states for periodical systems (see Ref. \cite{dod78}). The
construction of geometrical phase may be repeated also for
the nonperiodical quantum systems with imaginary characteristic times.
For systems with the imaginary characteristic times the generalized
loss--phase states may be constructed.

So, we answered the following question. Is the geometrical phase the
integral of motion of the nonstationary and nonperiodical quantum system?
The answer is "yes." The phase is the integral of motion. This phase is
the geometrical phase of the quantum system since the
spectrum of the operator $~U(T)$ and the spectrum of the operator
$~U(t)U(T)U^{-1}(t)$ coincide. Thus we have proved that the geometrical
phases of nonstationary systems are integrals of motion for these systems.
The introduced integrals of motion are of the same nature that have the
time--dependent integrals of motion for the harmonic oscillator with
varying frequency and a charge moving in varying magnetic field
constructed in Refs. \cite{mal70}, \cite{Malkinmankotrifonov69} and
\cite{tri70}. Such integrals of motion have been analyzed and applied to
general problems of quantum mechanics and statistics in Ref.
\cite{mal79},\cite{dod89}.

\section{Nonstationary Casimir Effect and Particle Distribution
Deformation}
The nonstationarity of Hamiltonian of a system produces some effects.
Among them nonstationary Casimir effect is of special interest
\cite{jslr}. It may exist in a resonator with moving walls.
The initial photon distribution is deformed due to this effect. So
the vacuum state becomes squeezed and correlated photon state
created by the external forces moving the walls of the resonator.
The state created may be described in terms of Wigner function.
Following \cite{olga} we will give the result for photon distribution
function of generic mixed squeezed and correlated Gaussian state.
Initially the state is taken to be standard coherent one (partial case
of such state is photon vacuum) and due to nonstationary Casimir effect
it becomes multimode mixed correlated state.

The most general mixed squeezed state of the $~N$--mode light with a
{\em Gaussian\/} density operator $~\hat{\varrho }$ is described by
the Wigner function $~W({\bf p},{\bf q})$ of the generic Gaussian form,
\begin{equation}
W({\bf p},{\bf q})=(\det {\bf M})^{-\frac 12}\exp\left
[-\frac 12({\bf Q}-<{\bf Q}>){\bf M}^{-1}({\bf Q}-<{\bf Q}>)\right],
\end{equation}
where $~2N$--dimensional vector $~{\bf Q}=({\bf p},{\bf q})$ consists
of $~N$ components $~p_1,...,p_N$ and $~N$ components $~q_1,...,q_N$,
operators $~\hat {{\bf p}}$ and $~\hat {{\bf q}}$ being the
quadrature components of the photon creation
$~\hat {{\bf a}}\dag$ and annihilation $~\hat {{\bf a}}$
operators (we use dimensionless variables and assume $~\hbar =1$):
\begin{eqnarray}
\hat {{\bf p}}&=&\frac {\hat {{\bf a}}
-\hat {{\bf a}}\dag}{i\sqrt {2}},\nonumber\\
\hat {{\bf q}}&=&\frac {\hat {{\bf a}}
+\hat {{\bf a}}\dag}{\sqrt {2}}.
\end{eqnarray}
$~2N$ parameters $~<p_i>$ and $~<q_i>$, $~i=1,2,\ldots ,N$, combined
into vector $~<{\bf Q}{\bf >}$, are the average values of the
quadratures,
\begin{eqnarray}
<{\bf p}>&=&\mbox{Tr}~\hat{\varrho}\hat {{\bf p}},\nonumber\\
<{\bf q}>&=&\mbox{Tr}~\hat{\varrho}\hat {{\bf q}}.
\end{eqnarray}
A real symmetric dispersion matrix $~{\bf M}$ consists of $~2N^2+N$
variances
\begin{equation}
{\cal M}_{\alpha\beta}=\frac 12\left\langle\hat Q_{\alpha}\hat
Q_{\beta}+\hat Q_{\beta}\hat Q_{\alpha}\right\rangle -\left\langle
\hat Q_{\alpha}\right\rangle\left\langle\hat Q_{\beta}\right\rangle
,~~~~~~~~~~~\alpha ,\beta =1,2,\ldots ,2N.
\end{equation}
\noindent They obey certain constraints, which are nothing but the
generalized uncertainty relations \cite{dod89}.

The photon distribution function of this state
\begin{equation}
{\cal P}_{{\bf n}}=\mbox{Tr}~\hat{\varrho }|{\bf n}><{\bf n}|,
{}~~~~~~~{\bf n}=(n_1,n_2,\ldots ,n_N),
\end{equation}
where the state $~|{\bf n}>$ is photon number state was calculated in
\cite{olga} and it is
\begin{equation}
{\cal P}_{{\bf n}}={\cal P}_0\frac {H_{{\bf n}{\bf n}}^{
\{{\bf R}\}}({\bf y})}{{\bf n}!},
\end{equation}
where
\begin{equation}
{\cal P}_0=\left[\det\left({\bf M}+\frac 12{\bf I}_{2N}\right)\right
]^{-\frac 12}\exp\left[-<{\bf Q}>\left(2{\bf M}+{\bf I}_{2N}\right
)^{-1}<{\bf Q}>\right],
\end{equation}
and
\begin{equation}
{\bf y}=2{\bf U}^t({\bf I}_{2N}-2{\bf M})^{-1}<{\bf Q}>,
\end{equation}
and the $2N$-dimensional unitary matrix
\begin{equation}
{\bf U}=\frac 1{\sqrt {2}}\left(\begin{array}{cc}
-i{\bf I}_N&i{\bf I}_N\\
{\bf I}_N&{\bf I}_N\end{array}\right)
\end{equation}
is introduced, in which $~{\bf I}_N$ is the $~N\times N$ identity
matrix. Also we use the notation
$${\bf n}!=n_1!n_2!...n_N!.$$
The function $~H_{{\bf n}{\bf n}}^{\{{\bf R}\}}({\bf y})$ is
multidimensional Hermite polynomial.

The mean photon number for j--th mode is expressed in terms of
photon quadrature means and dispersions
\begin{eqnarray}
<n_j>=\frac 12(\sigma_{p_jp_j}+\sigma_{q_jq_j}-1)
+\frac 12(<p_j>^2+<q_j^2>).
\end{eqnarray}
This formula corresponds to deformation of photon distribution due to
nonststionary Casimir effect.

\section{Q-nonlinearity and Planck Distribution}
Another reason to deform photon distribution function may be
related to q--nonlinearity \cite{sol1}, \cite{sol2} and below we review
this approach. Due to it the q--oscillators \cite{bie}, \cite{mc} may
be interpreted as nonlinear oscillators with special kind of the
nonlinearity. In \cite{sol1}, \cite{sol2} the influence of this
nonlinearity on the photon distribution function has been evaluated.
So the q--oscillator amplitude $~a_{q}$ is expressed in terms of usual
amplitude $~a$ obeying boson commutation relations
\begin{equation}
a_{q}=a\sqrt {\frac {\sinh \lambda a\dag a}{a\dag a \sinh \lambda }},~
{}~~~~~~~~~q=\exp \lambda.
\end{equation}
Then taking for the Hamiltonian of the classical q--oscillator the
expression
\begin{equation}
H=\frac {1}{2}(a_{q}\dag a_{q}+a_{q}a_{q}\dag ),
\end{equation}
and considering the nonlinear equation of motion for the amplitude of
photon $~a$ we have the solution of the equation of motion in the form
\begin{equation}
a(t)=a_{0}\exp [-i\omega (a_{0}\dag a_{0})t],~~~~~~~~\omega
=\frac {\lambda }{\sinh \lambda }\cosh \lambda a\dag a,
\end{equation}
which demonstrates the dependence of the frequency on the amplitude.
Calculating the partition function we have the following q--deformed
Planck distribution formula
\begin{equation}
<n>=\frac {1}{e^{\hbar \omega /kT}-1}-\lambda ^{2}\frac {\hbar \omega}
{kT}\frac {e^{3\hbar \omega /kT}+4e^{2\hbar \omega /kT}
+e^{\hbar \omega /kT}}{(e^{\hbar \omega /kT}-1)^{4}}.
\end{equation}
Thus we showed that nonstationarity (Casimir effect) and nonlinearity
deform the particle distribution finction.

\end{document}